\documentclass[prd, preprint, longbibliography, 11pt]{revtex4-1}

\usepackage{amsmath}
\usepackage{mathtools}
\usepackage{amssymb}
\usepackage{setspace}
\usepackage{graphicx}
\usepackage{natbib}
\usepackage{float}
\usepackage{tensor}
\usepackage[utf8]{inputenc}
\usepackage{amsfonts}
\usepackage{braket}
\usepackage{esint}
\usepackage{breqn}
\usepackage{IEEEtrantools}

\begin{document}
 
 %

\begin{center}
{ \large \bf Quantum Theory without Classical Time: Octonions, and a theoretical derivation of the Fine Structure Constant 1/137}

\smallskip

\vskip 0.1 in

{\large{\bf Tejinder P.  Singh }}

\smallskip

{\it Tata Institute of Fundamental Research,}
{\it Homi Bhabha Road, Mumbai 400005, India}\\
\smallskip
 {\tt tpsingh@tifr.res.in}

\end{center}
\vskip 1 in

\centerline{\bf ABSTRACT}
\smallskip

\noindent There must exist a reformulation of quantum field theory which does not refer to classical time. We propose a pre-quantum, pre-spacetime theory, which is a matrix-valued Lagrangian dynamics for gravity, Yang-Mills fields, and fermions. The definition of spin in this theory leads us to an eight dimensional octonionic space-time. The algebra of the octonions reveals the standard model; model parameters are determined by roots of the cubic characteristic equation of the exceptional Jordan algebra. We derive the asymptotic low energy value 1/137 of the fine structure constant, and predict the existence of universally interacting spin one Lorentz bosons, which replace the hypothesised graviton. Gravity is not to be quantized, but is an emergent four-dimensional classical phenomenon, precipitated by the spontaneous localisation of highly entangled fermions.
\vskip 1 in

\centerline{March 27, 2021}

\bigskip

\centerline{Essay written for the Gravity Research Foundation 2021 Awards for Essays on Gravitation}
\centerline{This essay received an Honorable Mention}
\centerline{Published Int. J. Mod. Phys. D 30 (2021) 2142010 DOI: 10.1142/S0218271821420104 }
\newpage

\bigskip

\noindent There ought to exist a reformulation of quantum [field] theory which does not depend on classical time. To arrive at such a reformulation, valid at the Planck scale, one constructs a  pre-quantum,  pre-spacetime theory, in two steps. In the first step, the pre-quantum theory is constructed by retaining classical space-time, and raising classical matter configuration variables and their corresponding canonical momenta to the status of matrices [equivalently, operators]. A matrix-valued Lagrangian dynamics is constructed, with the Lagrangian being the trace of a matrix polynomial. The matrices themselves have complex  Grassmann numbers as entries, with even-grade Grassmann matrices $q_B$(odd-grade matrices $q_F$) representing bosonic(fermionic) degrees of freedom. Matrix-valued equations of motion are derived from the trace Lagrangian, but the Heisenberg algebra is not imposed on the matrix commutators. Instead, the commutators evolve dynamically, and yet the global unitary invariance of the trace Hamiltonian implies  the existence of the following novel conserved charge, made from the commutators and anti-commutators:
\begin{equation}
\tilde{C} = \sum_i [q_{Bi}, p_{Bi}] - \sum_j \{q_{Fj}, p_{Fj}\}
\label{cons}
\end{equation} 
This is Adler's trace dynamics \cite{Adler:04}  assumed operational at the Planck scale, and the Hamiltonian of the theory is in general not self-adjoint. Next, we ask as to what is the emergent 
coarse-grained dynamics at low energies, if the trace dynamics is not being examined at Planck time resolution? If not too many degrees of freedom are entangled with each other, then the anti-self-adjoint part of the Hamiltonian is negligible, and the emergent dynamics is quantum [field] theory. The conserved charge mentioned above gets equipartitioned, and each commutator and anti-commutator in it is identified with $i\hbar$, where $\hbar$ is Planck's constant. On the other hand, if sufficiently many degrees of freedom are entangled, the anti-self-adjoint part of the Hamiltonian becomes significant, `collapse of the wave function' [spontaneous localisation] ensues, and ordinary classical dynamics is recovered. Qualitatively speaking, spontaneous localisation randomly maps each canonical matrix  to one or the other of its eigenvalues [Born probability rule is obeyed], and the conserved charge above goes to zero identically. Thus the underlying trace dynamics is a {\it deterministic  but non-unitary} pre-quantum theory [on a classical space-time background], and quantum field theory, along with its indeterminism,  is an emergent low-energy  thermodynamic phenomenon. The emergent indeterminism is a consequence of the coarse-graining: i.e.  not having precise information about the dynamics on  the small time-scales that have been averaged over.

In the second step, one goes from the above pre-quantum theory to a pre-quantum, pre-spacetime theory \cite{Singh2020DA}  by raising space-time points also to the status of matrices, via the following profound theorem \cite{Chams:1997}  in Riemannian geometry. The Einstein-Hilbert action is proportional to the trace of the regularised Dirac operator $D_B$ on the manifold, in a truncated heat-kernel expansion in $L_P^{-2}$: 
\begin{equation}
Tr \; [L_P^2 \; D_B^2] \sim \int d^4x \sqrt{g} \; \frac{R}{L_P^2} + {\cal O}(L_P^0)\sim L_P^2 \sum_n \lambda_n^2
\label{grraction}
\end{equation}
The eigenvalues $\lambda_n$ of the Dirac operator play the role of the dynamical variables of general relativity \cite{Rovelli}.
To go from here to the pre-quantum, pre-spacetime theory, every eigenvalue  $\lambda_n$ is raised to the status of a canonical matrix momentum: $\lambda_n\rightarrow p_{Bn}\propto q_{Bn}/d\tau \equiv D_B$, where the bosonic matrix $q_B$ is the corresponding newly introduced configuration variable.  We hence have $N$ copies of the Dirac operator ($n$ runs from 1 to $N$,  with  $N\rightarrow \infty$).  The matrix dynamics trace Lagrangian [space-time part] for the $n$-th degree of freedom is then $Tr \; (dq_{Bn}/d\tau)^2$ and the total matrix dynamics action [space-time part] is  $S \sim \sum_n \int d\tau\; Tr\; (dq_{Bn}/d\tau)^2$. Here, $\tau$ is an absolute time-parameter, known as Connes time, which is a unique feature of a {\it non-commutative} geometry [which is what we now have]  and is a consequence of the Tomita-Takesaki theory. Yang-Mills fields are represented by the matrices $q_{Bn}$, gravitation by the $\dot{q}_{Bn}$,  fermionic degrees of freedom by  fermionic matrices $q_{Fn}$ and their `velocities' $\dot{q}_{Fn}$, where `dot' denotes a derivative with respect to the time $c\tau$. 
Each of the $n$ degrees of freedom has an action of its own, which is given  by \cite{Singh2020DA}
\begin{equation}
\frac{S}{C_0} =  \frac{a_0}{2} \int \frac{d\tau}{\tau_{Pl}} \; Tr  \bigg[\dot{q}_B^{\dagger} + i\frac{\alpha}{L} q_B^\dagger+ a_0 \beta_1\left( \dot{q}_F^\dagger  + i\frac{\alpha}{L} q_F^\dagger\right)\bigg] \times \bigg[ \dot{q}_B + i\frac{\alpha}{L} q_B+ a_0 \beta_2\left( \dot{q}_F + i\frac{\alpha}{L} q_F\right)\bigg] 
\label{ymi}
\end{equation} 
where  $a_0 \equiv L_P^2 / L^2$. The total action of this generalised trace dynamics is the sum over $n$ of $N$ copies of the above action, one copy for each degree of freedom. This total action defines the pre-spacetime, pre-quantum theory, with each degree of freedom [defined by the above action] being an `atom' of space-time-matter [an STM atom, or an `aikyon']. In an aikyon, one loses the distinction between space-time and matter: the fermionic part $q_F$ (say an electron) is the source for the bosonic part $q_B$; however the interpretation can also be reversed: $q_B$ can be thought of as the source for $q_F$. In this action, there are only three fundamental constants: Planck length, Planck time, and the constant $C_0$ having dimensions of action, to be identified with Planck's constant $\hbar$ in the emergent theory. $L$ is a length parameter [scaled with respect to $L_P$; $q_B$ and $q_F$ have dimensions of length] characterising the STM atom, and $\alpha$ is the dimensionless Yang-Mills coupling constant. $\beta_1$ and $\beta_2$ are two unequal complex Grassmann numbers: their being equal gives inconsistent equations of motion. The aikyon is therefore a 2D object.

The further analysis of this pre-space-time, pre-quantum theory proceeds just as for the pre-quantum trace dynamics. Equations of motion are derived, and there is a conserved charge just as in Eqn. (\ref{cons}). Assuming that the theory holds at the Planck scale, the emergent low-energy approximation obeys quantum commutation rules, and  is the sought for reformulation of quantum theory without classical time. This emergent theory is also a quantum theory of gravity. If sufficiently many aikyons get entangled, the anti-self-adjoint part of the Hamiltonian becomes significant, spontaneous localisation results, and the fermionic part of the entangled STM atoms is localised. There emerges a 4D classical space-time manifold (labelled by the positions of collapsed fermions), sourced by point masses and gauge fields, and  whose geometry obeys the laws of general relativity; [space-time from collapse of the wave-function]. Those aikyons which are not sufficiently entangled remain quantum; their dynamics is described by quantum field theory on  space-time background generated by the entangled, collapsed fermions [the macroscopic bodies of the universe]. 

However, this is not the end of the story! The action (\ref{ymi}) packs a great deal of new information, as we now unravel. To begin with, since we have a fundamental  Lagrangian dynamics, we should be able to define spin [angular momentum canonical to some appropriate angle] for the bosonic and fermionic degrees of freedom, and prove the spin-statistics theorem. After defining new dynamical variables $\dot{\widetilde Q}_B$ and $\dot{\widetilde{Q}}_F$ as
\begin{equation}
{\dot{\widetilde{Q}}_B} \equiv \frac{1}{L} (i\alpha q_B + L \dot{q}_B); \qquad  {\dot{\widetilde{Q}}_F} \equiv \frac{1}{L} (i\alpha q_F + L \dot{q}_F)
\label{lilqu}
\end{equation}
the Lagrangian in (\ref{ymi}) can be brought to the elegant and revealing form, akin to  a free particle:
\begin{equation}
 \mathcal{L} = \frac{a_0}{2} \; Tr  \biggl(\biggr. \dot{\widetilde{Q}}_{B}^\dagger + \dfrac{L_{p}^{2}}{L^{2}} \beta_{1} \dot{\widetilde{Q}}_{F}^{\dagger} \biggl.\biggr) \biggl(\biggr. \dot{\widetilde{Q}}_{B} + \dfrac{L_{p}^{2}}{L^{2}} \beta_{2} \dot{\widetilde{Q}}_{F} \biggl.\biggr) \biggl.
\label{eq:tracelagn}
\end{equation}
Next we introduce self-adjoint bosonic operators $R_B$ and $\theta_B$, and self-adjoint fermionic operators $R_F$ and $\theta_F$, as follows:
$\widetilde{Q}_B \equiv R_B\; \exp i\theta_B \ ;  \  \widetilde{Q}_F \equiv R_F \; \exp i\eta \theta_F$.
Here, $\eta$ is a real Grassmann number, introduced to ensure that the fermionic phase is bosonic, so that $\widetilde{Q}_F$ comes out fermionic, as desired, upon the Taylor expansion of its phase. The canonical angular momenta corresponding to the angles $\theta_B$ and $\theta_F$ are bosonic and fermionic spin, which obeys the spin-statistics theorem, and agrees with our conventional understanding of spin in relativistic quantum mechanics \cite{Singh2020DA}.

But in which space  are these  angles $\theta_B$ and $\theta_F$ located? They cannot be in 4-D space-time because then the canonical momentum will be orbital angular momentum, not spin. We note from Eqn. (\ref{lilqu}) that the velocity $\dot{q}_B$ and the configuration variable $q_B$ together define a complex plane: spin measures the angular momentum of the periodic motion of $\widetilde{Q}_B$ in this plane. Making the plausible assumption that the velocity $\dot{q}_B$ has four components, which lie in our 4D space-time, we are compelled to conclude that the $q_B$ also has four components, which lie on four `internal' directions, thus making physical space eight dimensional! We already know that the $q_B$ describe Yang-Mills fields: so we have at hand a Kaluza-Klein type of theory, and spin describes periodic motion from 4D space-time to and from the internal directions. 

Furthermore, the coordinates in this 8D physical space will be assumed to be non-commuting,  consistent with the fact that the matrices $q_B$ and $\dot{q}_B$ do not commute. The introduction of an 8D non-commuting coordinate system immediately suggests the eight dimensional number system known as the octonions: the largest of the only four normed division algebras (reals, complex numbers, quaternions, octonions). We will assume that  the matrices in our theory are each made of eight-component Grassmanian matrices, one component for each of the eight directions of an octonionic space-time, where an octonion is the number
\begin{equation}
O = a_0 + a_1 {\bf e_1} + a_2 {\bf e_2} + a_3 {\bf e_3} + a_4 {\bf e_4} + a_5 {\bf e_5} + a_6 {\bf e_6} + a_7 {\bf e_7} 
\end{equation}
with $(a_0, a_1, ..., a_7)$ being real numbers, and each of the ${\bf e_i}$ is equal to $\sqrt{-1}$, and the seven ${\bf e_i}$ anti-commute and obey the Fano plane multiplication rules. The automorphisms of the octonions take over the role of space-time diffeomorphisms and internal gauge transformations, thus unifying the latter two. The [non-associative] algebra of the octonion automorphisms forms the smallest of the five exceptional Lie groups $G_2$, which has fourteen generators. The unitary transformations generated by these act on the $q$-matrices while leaving the trace Lagrangian invariant, and once again yield the conserved charge of Eqn. (\ref{cons}). An aikyon described by the Lagrangian (\ref{ymi}) evolves in this eight dimensional octonionic space-time in Connes time $\tau$ and by virtue of the isomorphism $SL(2, {\mathbb O}) \simeq SO(1,9)$ this is also a 10D Minkowski space-time. Add to this the Connes time as the eleventh dimension, and the 2D aikyon moving in this background is reminiscent of 11D M-theory, and by way of this pre-spacetime, pre-quantum theory we have a non-perturbative formulation of the dynamics, {\it ab\ initio}. Spontaneous localisation confines classical systems to a 4D Minkowski space-time obeying general relativity, whereas quantum systems reside in the full 8D space,  though they can be described from a 4D space-time perspective as well, provided we adopt the conventional definition of quantum spin.

The non-commutative, non-associative algebra of the octonions, in conjunction with the Lagrangian (\ref{ymi}), severely constrains and determines the allowed properties and quantum numbers of the bosonic and fermionic degrees of freedom.  The group $G_2$ has two intersecting maximal sub-groups: one the $U(3) \sim SU(3) \times U(1)$, this being the element preserver group of the imaginary octonions,  and the other $SO(4)\sim SU(2) \times SU(2)$, this being the stabiliser group of the quaternions inside the octonions. The two maximal sub-groups have an intersection $U(2)\sim SU(2) \times U(1)$ and between them account for the 14 generators of $G_2$, i.e. (8+6=14). A Clifford algebra $Cl(6)$ constructed from the element preserving set of [complexified] octonions  implies an eight dimensional fermionic basis of states, for which a convenient representation turns out to be as follows \cite{f3, Singh2020DA} 
\begin{equation}
\begin{split}
 &\frac{i}{2} e_7 \quad [{\rm Neutrino}]; \qquad  -\frac{1}{4}(i+e_7) \quad [{\rm  Positron}]\\
 &\frac{1}{4} ( e_5 + ie_4) \ ; \qquad
 \frac{1}{4} ( e_3 + ie_1) \ ; \qquad 
  \frac{1}{4} ( e_6 + ie_2)  \qquad \ [{\rm Anti-down\ quarks}] \\
&\frac{1}{4} ( e_4+ ie_5)\ ; \qquad
 \frac{1}{4} ( e_1 + ie_3) \ ; \qquad
 \frac{1}{4} ( e_2 + ie_6) \qquad [{\rm Up\ quarks}]
 \label{firstgen}
\end{split}
\end{equation}
The two states in the first row transform as singlets under $SU(3)$, those in the second row as anti-triplets, and those in the third row as triplets. A number operator $U(1)/3$ constructed from the $U(1)$ above has eigenvalues 0 and 1 for the first two states, eigenvalues 1/3 for every state in the second row, and 2/3 for the states in the third row. These results suggest interpreting the eigenvalues as electric charge, and imply the particle interpretation shown against the states, and anti-particles are complex conjugates of these states. Thus the algebra implies there are eight fermions [two leptons, six quarks] and their eight anti-particles per generation. The $SU(3)\times U(1)$ is identified with the color-electro symmetry $SU(3)_c\times U(1)_{em}$.  

Alternatively, a $Cl(6)$ can be constructed from the stabiliser group $SU(2) \times SU(2)$ and its action on the fermion basis states corresponds to that of the weak symmetry {\it and} the Lorentz symmetry, We have a Lorentz-Weak unification, with gravitation emerging only as a classical condensate from spontaneous localisation of many entangled aikyons. However it is not possible to simultaneously get two copies of $Cl(6)$  from the octonion algebra. One is compelled to consider the next number system in the Cayley-Dickson construction, the sedenions, whose automorphism group is equal to $Aut({\mathbb O})\times S_3$, and the isomorphism of the permutation group $S_3$ to Spin(8), known for its triality, suggests three fermion generations, described by three intersecting copies of $G_2$, and the Clifford algebra $Cl(14)$. The three copies of $G_2$ are embedded in the next larger Lie group $F_4$, which is also the automorphism group of the Exceptional Jordan Algebra $J_3({\mathbb O})$ of 3x3 Hermitean matrices with octonionic entries.  The bosons lie in the intersection, the gauge symmetry is color-electro-weak-Lorentz; and the 14 $G_2$  generators give rise to 14 spin one gauge bosons: 8 gluons, 4 electro-weak bosons, and the two newly predicted Lorentz bosons, which should be looked for in experiments. This interpretation of the particle content is fully supported by the Lagrangian (\ref{ymi}) which can be written as follows, and shown to describe 14 gauge bosons, the Higgs, and three fermion generations \cite{Singhfsc}:
\begin{equation}
\begin{split}
&{\cal L} \equiv \frac{L_P^2}{2L^2} \;  Tr \left[ {\cal L}_{b} + {\cal L}_{1} + {\cal  L}_{2} + {\cal L}_{3} +{\cal L}_{4} \right]\ ; \quad
{\cal L}_{b} =  \dot{q}_{B}^\dagger \dot{q}_{B}  -\frac{\alpha^2}{L^2}  q_B^\dagger {q}_B +\frac{i\alpha}{L}  q_B^\dagger \dot{q}_B + \frac{i\alpha}{L}  \dot{q}_{B}^\dagger {q}_{B}
\ ; \\
&{\cal L}_{1} =  \frac{L_P^2}{L^2} \dot{q}_B^\dagger \beta_2 \dot{q}_F + \frac{L_P^2}{L^2} \beta_1 \dot{q}_F^\dagger \dot{q}_B +  \frac{L_P^2}{L^2} \dot{q}_B \beta_2 \dot{q}_F\ ;
\quad
{\cal L}_{2} = -\frac{\alpha^2}{L^2} \left(\frac{L_P^2}{L^2} q_B^\dagger \beta_2 {q}_F + \frac{L_P^2}{L^2} \beta_1 q_F^\dagger {q}_B +\frac{L_P^2}{L^2} {q}_B \beta_2 {q}_F\right) ;
\\
&{\cal L}_{3} = \frac{i\alpha}{L} \left(  \frac{L_P^2}{L^2} q_B^\dagger \beta_2 \dot{q}_F + \frac{L_P^2}{L^2} \beta_1 q_F^\dagger \dot{q}_B +  \frac{L_P^2}{L^2} {q}_B^\dagger \beta_1 \dot{q}_F^\dagger  \right) \ ;
\ \ 
{\cal L}_{4} =  \frac{i\alpha}{L} \left(\frac{L_P^2}{L^2} \dot{q}_B^\dagger \beta_2 {q}_F + \frac{L_P^2}{L^2} \beta_1 \dot{q}_F^\dagger {q}_B +   \frac{L_P^2}{L^2}\dot{q}_B^\dagger \beta_1 q_F^\dagger\right)
\end{split}
\end{equation}
Here, $({\cal L}_{b} , {\cal L}_{1} , {\cal  L}_{2} ,  {\cal L}_{3} , {\cal L}_{4})$ respectively describe the 14 gauge bosons, Lorentz-weak action on all fermions, photon action on charged fermions, gluon action on 1/3 charge quarks, and gluon action on 2/3 charge quarks. The coefficient $\alpha^2 (L_P/L)^4$ of ${\cal L}_2$ is the asymptotic fine structure constant, and in this theory  it is determined  by the roots of the cubic characteristic equation of $J_3({\cal O)}$ \cite{Singhfsc}:
\begin{equation}
\lambda^3 - Tr(X) \lambda^2 + S(X) \lambda - Det(X)=0
\label{cbe}
\end{equation}
where $Tr(X), S(X), Det(X)$ are  respectively the trace, a two form, and determinant of the 3x3 octonion valued Hermitean matrix $X$; the diagonal entries are assumed to be electric charge values: 
\begin{equation}
\begin{aligned}
& X(\xi, x)=
\begin{bmatrix}
\xi_1 &  x_3 & x_2^*\\
x_3^* & \xi_2 & x_1\\
x_2 & x_1^* & \xi_3
\end{bmatrix}\ ;
\ Det (X) = \xi_1 \xi_2 \xi_3 + 2 Re (x_1 x_2 x_3) - \sum_1^3 \xi_i x_i x_i^* \\
&\qquad \qquad \qquad \qquad  \qquad  \quad S(X) = \xi_1 \xi_2 - x_3 x_3^* + \xi_2 \xi_3 - x_1 x_1^* + \xi_1 \xi_3 - x_2^* x_2
\label{mat}
\end{aligned}
\end{equation}
The octonionic representation for second and third fermion generations can be calculated by appropriate rotations of the first generation (\ref{firstgen}), and then used in (\ref{mat}). Four sets of three eigenvalues each are calculated from this cubic equation, for  the four different values of electric charge of the fermions,  and are as follows: Neutrinos: $(-\sqrt{3}/2, 0, \sqrt{3}/2)$, the 1/3 quarks: $(1/3 -\sqrt{3/8}, 1/3, 1/3+\sqrt{3/8})$, the 2/3 quarks: $(2/3 -\sqrt{3/8}, 2/3, 2/3+\sqrt{3/8})$, and the charged leptons: $(1 -\sqrt{3/8}, 1, 1+\sqrt{3/8})$. The octonionic magnitudes for these four fermion sets are respectively $\sum_1^3  x_ix^i = \sqrt{3}/2, \sqrt{3/8}, \sqrt{3/8}, \sqrt{3/8}$ and these are set equal to the scale $(2L_P/L)$ of the corresponding fermion. The constant $\alpha$ is fixed by demanding the exponential variation $d\alpha /dq \propto \alpha$, where $q$ is the electric charge, and then determining $\alpha$ from the smallest electric charge = 1/3 for the anti-down quark, with the proportionality constant being the anti-down quark eigenvalue $(1/3 - \sqrt{3/8})$. Taking $L_P/L$ to be $\sqrt{3/32}$ for the charged leptons, as just mentioned, the low-energy fine structure constant is deduced to have the value \cite{Singhfsc}
\begin{equation}
\alpha^2 \left(\frac{ L_P} { L}\right) ^4\equiv e^2 /\hbar c =  \exp \left[ \left [ \frac{1}{3} - \sqrt\frac{3}{8} \right ] \times \frac{2}{3} \right] \times \frac{9}{1024} \approx  0.00729713 = \frac{1}{137.04006}
\label{fsc}
\end{equation}
which agrees with the measured value to 2 parts in $10^7$, and agrees with it  exactly if the Karolyhazy correction is accounted for, and a specific energy scale for the electro-weak symmetry breaking is assumed \cite{Singhfsc}. The electric charge $e$ of the electron is hence derived in terms of the three fundamental constants $\hbar, L_P, \tau_{Pl}$ of our theory, and is not a new independent parameter.
Additional significant  eigenvalues of $J_3({\mathbb O})$ can  be calculated as sets of three each, for each of the three generations of charged fermions, and are given by $[ \ 2/3+2\sqrt{-Q} \cos \theta_i/3, \ 2/3 + 2\sqrt{-Q} \cos (\theta_i +2\pi)/3, \ 2/3 + 2\sqrt{-Q} \cos (\theta_i+4\pi)/3 \ ]$. Here, $Q=-35/216$, $i$ takes the value $I, II, III$ for  the three generations, and the angles are found to be $\theta_I = 1.81270, \theta_{II}= 1.69730, \theta_{III} = 1.74837$. The eigenvalues correlate with mass ratios \cite{Singhfsc,vatsalya1,vvs,Singhreview} and the fact that $\theta_I \neq \theta_{II} \neq \theta_{III}$ suggests violation of lepton universality.

We began by addressing a quantum foundational issue: to remove classical time from quantum theory. The answer leads us to a candidate theory of quantum gravity and unification, and a theoretical derivation of the fine structure constant 1/137. And we find that the graviton is replaced by two spin one Lorentz bosons [a verifiable prediction] on par with the other gauge bosons.


\bibliography{biblioqmtstorsion}

\end{document}